\documentclass[12pt,a4paper,final]{iopart}

\usepackage{iopams}
\usepackage{graphicx}
\usepackage[breaklinks=true,colorlinks=true,linkcolor=blue,urlcolor=blue,citecolor=blue]{hyperref}
\usepackage{mathrsfs}
\def\tr{{\rm tr}}

\newtheorem{proposition}{Proposition}

\begin{document}

\title[Feedback Policies  for Measurement-based Quantum State Manipulation]{Feedback Policies  for Measurement-based Quantum State Manipulation}

\author{Shuangshuang Fu}
\address{College of Engineering and Computer Science}
\address{The Australian National University, Canberra, Australia}
\ead{shuangshuang.fu@anu.edu.au}

\author{Guodong Shi}
\address{College of Engineering and Computer Science}
\address{The Australian National University, Canberra, Australia}
\ead{guodong.shi@anu.edu.au}

\author{Alexandre Proutiere}
\address{School of Electrical Engineering}
\address{Royal Institute of Technology, Stockholm, Sweden}
\ead{alepro@kth.se}

\author{Matthew R. James}
\address{College of Engineering and Computer Science}
\address{The Australian National University, Canberra, Australia}
\ead{matthew.james@anu.edu.au}

\begin{abstract}
In this paper, we propose  feedback designs  for  manipulating a quantum state to a target state by performing sequential   measurements. In light of  Belavkin's quantum feedback control theory, for a given set of (projective or non-projective) measurements and a given time horizon, we show that finding the measurement selection  policy  that   maximizes  the probability of successful state manipulation is an optimal control problem for a  controlled Markovian process. The optimal policy is Markovian  and  can be solved by dynamical programming. Numerical examples  indicate that making use of  feedback information  significantly improves the success probability compared to classical scheme without taking feedback. We also consider other objective functionals including maximizing the expected fidelity to the target state as well as minimizing the expected arrival time. The connections and differences among these objectives are also discussed.
\end{abstract}

\vspace{2pc}

\section{Introduction}

One fundamental difference between classical and quantum mechanics is the unavoidable back-action of quantum measurement. On the one hand, this back-action is generally thought to be detrimental for the implementation of effective quantum control. On the other hand, it also provides us one possibility to use the change caused by the measurement as a new route to manipulate the state of the system\cite{belavkin1983towards,wiseman1993quantum}. A basic problem in quantum physics and engineering is how to drive a quantum  system to a desired target state. There have been studies on the preparation of a given target state  from a given initial state  using sequential (projective or non-projective) measurements in the last few years \cite{ashhab2010control,jacobs2010feedback,wiseman2011quantum,roa2007quantum,pechen2006quantum}.

A quantum measurement $E$ is described by a collection of measurement  operators $$
\Big\{\mathsf{M}_E(m) \Big\}_{m\in \mathcal{Y}},
$$
 where $\mathcal{Y}$ is an index set for measurement outcomes and the measurement operators satisfy
$$
\sum_{m\in \mathcal{Y}} \mathsf{M}_E(m)^\dag \mathsf{M}_E(m)=I.
$$
Suppose we perform the quantum measurement $E$ on density operator $\rho$, the probability of obtaining result $m\in\mathcal{Y}$ is $\tr(\mathsf{M}_E(m) \rho \mathsf{M}_E(m)^\dag)$, and when  $m\in\mathcal{Y}$ occurs,  the post-measurement state of the quantum system becomes
$$
\mathcal{M}_{_E}^m(\rho)=\frac{\mathsf{M}_E(m) \rho \mathsf{M}_E(m)^\dag}{\tr(\mathsf{M}_E(m) \rho \mathsf{M}_E(m)^\dag)}.
$$

If we are unaware of the measurement result, the unconditional state of the quantum system after the measurement can be expressed as
$$
\mathcal{M}_{E}(\rho)=\sum_{m\in \mathcal{Y}}\mathsf{M}_E(m) \rho \mathsf{M}_E(m)^\dag.
$$
If $\{\mathsf{M}_E(m)\}_{m\in \mathcal{Y}}$ are orthogonal projectors, i.e., the $\mathsf{M}_E(m) $ are Hermitian and $\mathsf{M}_E(l)\mathsf{M}_E(m)=\delta_{lm}\mathsf{M}_E(m)$, $E$ is a projective measurement. The idea of quantum state manipulation using sequential measurements \cite{ashhab2010control,jacobs2010feedback,wiseman2011quantum,roa2007quantum,pechen2006quantum} is as follows. By consecutively performing the measurements $E_1,\dots, E_N$, the unconditional state for quantum system with initial state $\rho_0$ can be expressed as
$$
\rho_{_N}^{\rm u}=\mathcal{M}_{E_N} \circ \mathcal{M}_{E_{N-1}} \circ \cdots \circ \mathcal{M}_{E_{1}}(\rho_0).
$$
It has been shown, analytically or numerically, how to select the measurements  $E_1,\dots, E_N$ so that $\rho_{_N}^{\rm u}$ can asymptotically tend to a desired target state \cite{ashhab2010control,jacobs2010feedback,wiseman2011quantum,roa2007quantum,pechen2006quantum}.

 Making use of feedback information for quantum measurement and detection actually has a long history, which can be viewed as the dual problem of state manipulation.  The ``Dolinar's receiver" proposes a feedback strategy for discriminating  two possible quantum states with prior distribution with minimum probability of error \cite{Dolinar1973}. The problem is known as the quantum detection problem and Helstrom's bound characterizes the minimum probability of error for discriminating any two non-orthonormal states \cite{Helstrom}. Quantum detection is to identify uncertain quantum states via projective measurements; while the considered quantum state projection is to manipulate a certain quantum state to a certain target, again via projective measurements. The Dolinar's scheme follows a similar structure that measurement is selected based on previous measurement results on different segments of the pulse, and was recently realized experimentally \cite{Cook2007}.  See \cite{WisemanIEEESurvey2009} for a survey for the extensive studies in feedback (adaptive) design in quantum tomography.

In this paper, we propose a feedback design for  quantum state manipulation via sequential measurements.  For a given set of measurements and a given time horizon, we show that finding the policy of measurement selections that   maximizes  the probability of successful state manipulation can be solved by dynamical programming. Such derivation of the optimal policy falls to  Belavkin's  quantum feedback theory \cite{belavkin1983towards}. Numerical examples are given which indicate that the proposed feedback policy significantly improves the success probability compared to  classical policy  by consecutive projections without taking feedback. In particular, the probability of reaching the target state $|1\rangle$ via feedback policy reaches  $0.9968$ using merely $10$ measurements from initial state $|0\rangle$. Other optimality criteria are also discussed such as the maximal expected fidelity and the minimal arrival time, and some connections and differences among the  the different criteria are also discussed.

The remainder of the paper is organised as follows. In the first part of Section \ref{sec2}, we  revisit a simple example of  reaching $|1\rangle$ from  $|0\rangle$ using sequential projective measurements \cite{pechen2006quantum},  and show  how feedback policies  work under which  even a little bit of  feedback can make a nontrivial  improvement. The rest of Section \ref{sec2} devotes to a rigorous treatment for the problem definition and for  finding the optimal feedback policy from classical quantum feedback theory. Numerical examples are given there. Section \ref{sec3} investigates some other optimality criteria and finally Section \ref{sec4} concludes the paper.

\section{Quantum State Manipulation by Feedback}\label{sec2}
\subsection{A Simple Example: Why Feedback?}
 Consider now a qubit system, i.e., a two-dimensional Hilbert space.   The initial state of the quantum system is  $|0\rangle\langle 0|$, and the target state is   $|1\rangle\langle 1|$. Given  $T \geq 2$   projective measurements from the set
\begin{eqnarray}\label{MeasurementSet}
\mathcal{E}=\Big\{E_i,\ \  i=1,2,\ldots, T \Big\}.
\end{eqnarray}
where $E_i=\big\{|\phi_i\rangle\langle \phi_i|,|\psi_i\rangle\langle \psi_i|\big\}$ with
$$|\phi_i\rangle=\cos\Big(\frac{\pi i}{2T}\Big)|0\rangle + \sin\Big(\frac{\pi i}{2T}\Big) |1\rangle
$$
and
$$
|\psi_i\rangle=-\sin\Big(\frac{\pi i}{2T}\Big)|0\rangle + \cos\Big(\frac{\pi i}{2T}\Big) |1\rangle.
$$
Note that the choice of $E_i$ follows the optimal selection given in \cite{pechen2006quantum}.

The strategy in \cite{roa2007quantum,pechen2006quantum} is simply to   perform the $T$ measurements in turn from $E_1$ to $E_T$.  We call it a {\it naive} policy. The probability of successfully driving the state from $|0\rangle $ to $|1\rangle$ in $T$ steps under this naive strategy is denoted by $p(T)$. We can easily calculate that $p(3)\approx0.56$ and $p(10)\approx0.8$.

Let $T=3$.   We next show that even only a bit of  measurement feedback can improve the performance of the strategy significantly.

\noindent{\it S1.  After the first measurement $E_1$ has been made, perform $E_3$ if the outcome is $|\psi_1\rangle$ for the second step, and follow the naive policy for all other actions.}

Following this scheme, it turns out that  the probability of arriving at $|1\rangle$ in three steps becomes around  $0.66$, in contrast with $p(3)\approx0.56$ under the naive scheme. The improvement in the probability of success comes from the fact that a feedback decision is made based on the information of the outcome of $E_1$ so that in {\it S1} a better selection of measurement is obtained between $E_2$ and $E_3$.

\subsection{Optimal Policy from Quantum Feedback Control}

We now present the solution to the optimal policy for  the considered quantum state manipulation in light of the classical work of quantum feedback control theory derived by Belavkin \cite{belavkin1983towards} (also see \cite{MattPRA} and \cite{MattSIAM-Review} for a thorough treatment).

Consider a quantum system whose state is described by density operators over the qubit space. Let $\mathcal{E}$ be a given finite set of measurements serving as all feasible control actions. For each $E\in \mathcal{E}$, we write
 $$
E=\Big\{\mathsf{M}_E(y)\Big\}_{y\in \mathcal{Y}},
 $$
where  $\mathcal{Y}$ is a finite index set of measurement outputs and $\mathsf{M}_E(y)$ is the measurement operator corresponding to outcome $y\in \mathcal{Y}$. Time is slotted with a horizon $N\geq 1$. The initial state of the quantum system is $\rho_0$,  and the target state is assumed to be, for the ease of presentation,  $|1\rangle\langle 1|$.

For $ 0\leq k \leq N-1$, we denote by $u_k\in \mathcal{E}$ the measurement performed at time $k$ and the post-measurement  state after $u_k$ has been performed is denoted as $\rho_{k+1}$. Let $y_k\in \mathcal{Y}$ be the outcome of $u_k$. The measurement sequence $\{u_k\}_{k=0}^{N-1}$ is selected by a {\it policy}  $\pi=\{\pi_k\}_{k=0}^{N-1}$, where each $\pi_s$ takes  value in the set $\mathcal{E}$  such  that  $u_{k}=\pi_k(y_0,\dots, y_{k-1}; u_0,\dots,u_{k-1})$ can depend on all previous selected measurements and their outcomes for all $k=0,\dots,N-1$. Here for convenience we have denoted $u_{-1}=y_{-1}=\emptyset$.

We can now express the closed-loop evolution of $\{\rho_k\}_0^N$ as
\begin{equation}\label{10}
\rho_{k+1}=\mathcal{M}_{u_k}^{y_k}(\rho_{k})=\frac{\mathsf{M}_{u_k}(y_k)\rho_k  \mathsf{M}_{u_k}^\dag(y_k) }{{\rm tr}\Big(\mathsf{M}_{u_k}(y_k)\rho_k  \mathsf{M}_{u_k}^\dag(y_k)\Big)},
\end{equation}
where $k=0,\dots,N-1$.
The distribution of $y_k$ is given by
$$
\mathbb{P}\Big(y_k=y\in\mathcal{Y} \Big| u_k, \rho_k\Big)={\rm tr}\Big(\mathsf{M}_{u_k}(y)\rho_k  \mathsf{M}_{u_k}^\dag(y)\Big),
$$
where $k=0,\dots,N-1$.
Clearly $\{\rho_k\}_0^N$ defines a Markov chain.

We define\footnote{ It is clear from this objective that $E_\ast=\{|0\rangle \langle 0|, |1\rangle \langle 1| \}$ must be a measurement in the set $\mathcal{E}$ for $\mathrm{J}_\pi (N)$ to be a non-trivial function if all measurements in $\mathcal{E}$ are projective.}
$$
\mathrm{J}_\pi (N):=\mathbb{P}_{\pi}\Big(\rho_{_N}=|1\rangle\langle 1| \Big)
$$
as the probability of successfully manipulating the quantum state to  the target density matrix $|1\rangle\langle 1|$, where  $\mathbb{P}_{\pi}$ is the probability measure equipped with $\pi$. We also define  the cost-to-go function
$$
\mathbf{V}(t,x)=\max_\pi \mathbb{P}\Big(\rho_{_N}=|1\rangle\langle 1|\Big|\rho_{_{N-t}}=x\Big)
$$
for $t=0,1,\dots,N$.
Following standard theories for controlled Markovian process \cite{Bertsekas,MDP}, the following conclusion holds.
\begin{proposition}
The cost-to-go function $\mathbf{V}(t,x)$ satisfies the following recursion
\begin{equation}
\mathbf{V}(t,x)= \max_{u\in\mathcal{E}}\sum_{y\in \mathcal{Y}}\mathbb{P}\Big(y\Big|u,x\Big)\mathbf{V}\Big(t-1, \mathcal{M}_{u}^{y}(x)\Big),
\end{equation}
where $t=1,\dots,N$,
with boundary condition
$\mathbf{V}(0,x)= 1$  if  $x=|1\rangle\langle 1|$,  and
  $\mathbf{V}(0,x)= 0$ otherwise.
The maximum arrival probability $\max_{\pi} \mathrm{J}_\pi (N)$ is given by
$\max_{\pi} \mathrm{J}_\pi (N)=\mathbf{V}(N,\rho_0)$. The optimal policy $\pi^\star=\{\pi_k^\star\}_{k=0}^{N-1}$ is Markovian, and is given by
\begin{equation}\label{1}
\pi^\star_k(\rho_{k})= \arg \max_{u\in\mathcal{E}}\sum_{y\in \mathcal{Y}}\mathbb{P}\Big(y\Big| u,\rho_{k}\Big)\mathbf{V}\Big(N-k-1, \mathcal{M}_{u}^{y}(\rho_{k})\Big)
\end{equation}
for $k=0,\dots,N-1$.
\end{proposition}

\subsection{Numerical Examples}

 We now compare the performance of the policies with and without feedback. Again we consider driving  a two-level  quantum system  from state  $|0\rangle$ to $|1\rangle$. The available measurements are in the set
$$
\mathcal{E}=\Big\{E_i,\ \  i=1,2,\ldots, T \Big\}.
$$
as given in Eq.(\ref{MeasurementSet}).

\subsubsection{Feedback vs. Non-Feedback}
\begin{figure}[t]
\begin{center}
\includegraphics[width=10cm]{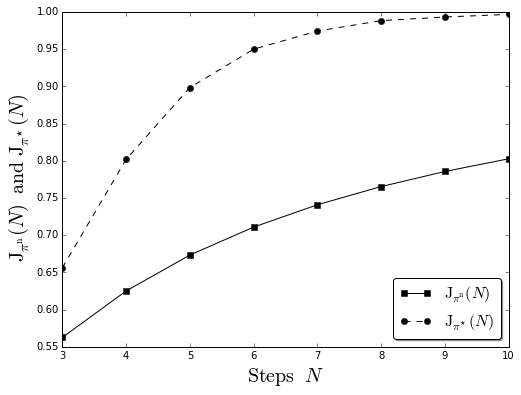}
\caption{The probabilities  of successfully reaching $|1\rangle$ from the initial state $|0\rangle$ using naive policy $\pi^{{\rm n}}$  and optimal feedback  policy $\pi^\star$, respectively. }
\label{success}
\end{center}
\end{figure}

First of all, we take $T=N$. The naive policy in turn takes projections from $E_1$ to $E_N$, denoted $\pi^{{\rm n}}=\{\pi^{{\rm n}}_k\}_{k=0}^{N-1}$. We solve the optimal feedback policy $\pi^\star=\{\pi_k^\star\}_{k=0}^{N-1}$ using Eq. (\ref{1}). It is clear that $\pi^{{\rm n}}$  is deterministic with $\pi^{{\rm n}}_k=E_{k+1}$, while $\pi^\star$ is Markovian with  $\pi_k^\star$ depending  on $\rho_{k}$. Correspondingly, their arrival probability in $N$ steps are given by $\mathrm{J}_{\pi^{\rm n}} (N)$ and $\mathrm{J}_{\pi^{\star}} (N)$, respectively.  In Figure \ref{success}, we plot  $\mathrm{J}_{\pi^{\rm n}} (N)$ and $\mathrm{J}_{\pi^{\star}} (N)$  for $N=3,\ldots,10$.  As shown clearly in the figure, the probability of success is improved significantly. Actually for $N=10$, we already have $\mathrm{J}_{\pi^{\star}} (N)=0.9968$.

Moreover, as an illustration of the different actions between the  naive and feedback  strategies, we plot their policies for $N=5$ in Tables I and II, respectively.

\begin{table}
\begin{center}
\renewcommand\arraystretch{1.0}
\resizebox{10cm}{!} {
\begin{tabular}{|c | c | c |c |c|c|}
 \hline
$\pi^{\rm n}$ & $k=0$ & $k=1$ & $k=2$ &$k=3$ & $k=4$\ \\
 \hline
$|0\rangle$ & $E_1$ & $*$ & $*$ & $*$& $*$\\
 \hline
$|1\rangle$ & $*$ & $*$ & $*$ & $*$ & $*$ \\
 \hline
$|\phi_1\rangle$ & $*$ & $E_2$ &$*$ &$*$ &$*$\\
 \hline
$|\psi_1\rangle$ & $*$ & $E_2$ &$*$ &$*$ &$*$\\
 \hline
 $|\phi_2\rangle$ & $*$ & $*$ &$E_3$&$*$ &$*$\\
 \hline
$|\psi_2\rangle$ & $*$ & $*$ & $E_3$ & $*$ & $*$\\
\hline
$|\phi_3\rangle$ & $*$ & $*$ &$*$&$E_4$ &$*$\\
 \hline
$|\psi_3\rangle$ & $*$ & $*$& $*$& $E_4$& $*$\\
\hline
$|\phi_4\rangle$ & $*$ & $*$ &$*$&$*$ &$E_5$\\
 \hline
$|\psi_4\rangle$ & $*$ & $*$& $*$& $*$& $E_5$\\
\hline
\end{tabular}
}
\end{center}
\caption{\rm The actions  using naive strategy $\pi^{{\rm n}}$  to prepare the target state $|1\rangle$, starting from $|0\rangle$, for $N=5$. Here  $E_i$ represents the  measurement that the policy chooses, and $*$ means that it is not possible to be in that state at the corresponding step.}
\end{table}

\begin{table}
\begin{center}
\renewcommand\arraystretch{1.0}
\resizebox{10cm}{!} {
\begin{tabular}{|c | c | c |c |c|c|}
 \hline
$\pi^\star$ & $k=0$ & $k=1$ & $k=2$ &$k=3$ & $k=4$\ \\
 \hline
$|0\rangle$ & $E_2$ & $E_2$ & $E_3$ & $E_3$&$ E_5$\\
 \hline
$|1\rangle$ & $E_5$ & $E_5$ & $E_5$ & $E_5$ & $E_5$ \\
\hline
$|\phi_1\rangle$ & $E_3$ & $E_3$ &$E_3$ &$E_3$ &$E_5$\\
 \hline
$|\psi_1\rangle$ & $E_5$ & $E_5$ &$E_5$ &$E_5$ &$E_5$\\
 \hline
 $|\phi_2\rangle$ & $E_4$ & $E_4$ &$E_3$&$E_3$ &$E_5$\\
 \hline
$|\psi_2\rangle$ & $E_1$  & $E_1$& $E_1$& $E_1$& $E_5$\\
 \hline
$|\phi_3\rangle$ & $E_4$ & $E_4$ &$E_4$&$E_4$ &$E_5$\\
 \hline
$|\psi_3\rangle$ &$E_1$ & $E_1$& $E_2$& $E_2$& $E_5$\\
\hline
$|\phi_4\rangle$ & $E_5$ & $E_5$ &$E_5$&$E_5$ &$E_5$\\
 \hline
$|\psi_4\rangle$ & $E_2$ & $E_2$& $E_2$& $E_2$& $E_5$\\
 \hline
\end{tabular}
}
\end{center}
\caption{\rm The actions  using optimal feedback policy $\pi^\star$  to prepare the target state $|1\rangle$ for $N=5$.}
\end{table}

\subsubsection{Influence of Measurement Set}
We now investigate how the size of the available measurement set $\mathcal{E}$ influences the successful arrival probability in $N$ steps under optimal feedback. In this case, the optimal arrival probability $\mathrm{J}_{\pi^{\star}} (N)$ is also a function of $T$, and we therefore rewrite $\mathrm{J}_{\pi^{\star}} (N)=\mathrm{J}_{\pi^{\star}}^T(N)$.

\begin{figure}[htbp]
\begin{center}
\includegraphics[width=10cm]{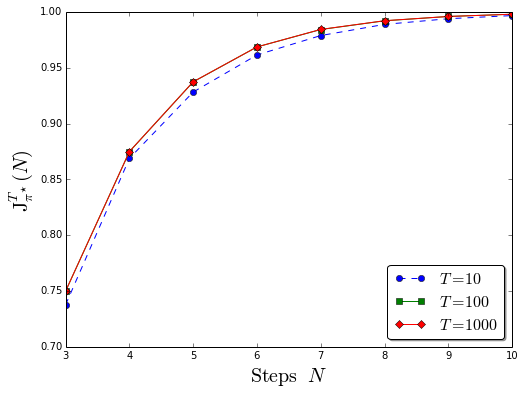}
\caption{The probabilities  of successfully reaching $|1\rangle$ from the initial state $|0\rangle$ using different sizes of measurement set by feedback strategy. }
\label{enlarge}
\end{center}
\end{figure}

In Figure \ref{enlarge}, we plot $\mathrm{J}_{\pi^{\star}}^T(N)$, for $T=10,100,1000$, respectively. The numerical results  show  that as $T$ increases, the $\mathrm{J}_{\pi^{\star}}^T(N)$ quickly tends to a limiting curve, suggesting the existence of  some fundamental upper bound on the arrival probability in $N$ steps using sequential projections from an arbitrarily large  measurement set.

\section{More Optimality Criteria}\label{sec3}

In this section, we discuss  two other  useful  optimality criteria, to maximize the expected fidelity with the target state, or to minimize the expected time it takes to arrive at the target state.
\subsection{Maximal  Expected Fidelity}
Given two density operators $\rho$ and $\sigma$, their {\it fidelity} is defined by \cite{Nielsen}
$$
F(\rho,\sigma)=\tr\sqrt{\sqrt{\rho}\sigma \sqrt{\rho}}.
$$
Fidelity measures the closeness of  two quantum states. Now that our target state $|1\rangle \langle 1|$ is a pure state, we have
$$
\tr\sqrt{\sqrt{|1\rangle \langle 1|}\sigma \sqrt{|1\rangle \langle 1|} }= \sqrt{\langle 1 | \sigma | 1 \rangle}.
$$

Alternatively, we can consider the following objective functional
$$
\tilde{\mathrm{J}}_\pi (N)=\mathbb{E}_\pi \Big[\langle 1| \rho_{_N} |1\rangle \Big],
$$
and the goal is to find a  policy that maximizes $\tilde{\mathrm{J}}_\pi (N)$.

For the two objective functionals  ${\mathrm{J}}_\pi (N)$ and $\tilde{\mathrm{J}}_\pi (N)$, we denote their corresponding optimal policy as $\pi^{\star }(N)=\{\pi^{\star}_k(N)\}_{k=0}^{N-1}$ and $\pi^\diamondsuit(N)=\{\pi^{\diamondsuit}_k(N)\}_{k=0}^{N-1}$, respectively, where the time horizon $N$ is also indicated.

Let $\pi^\diamondsuit(N-1)\oplus E_\ast$ be the policy that follows  $\pi^\diamondsuit(N-1)$ for $k=0,\dots, N-2$ and takes value $E_\ast$ for $k=N-1$. Let $\rho^{\rm u}_k$ be the unconditional  density operator at step $k$ for $k=0,\dots,N-1$. The following equations hold:
\begin{eqnarray}
\tilde{\mathrm{J}}_\pi (N-1)&=\mathbb{E}_\pi \Big[\langle 1| \rho_{_{N-1}} |1\rangle \Big]\nonumber\\
 &=\tr \Big ( \rho_{_{N-1}}^{\rm u} |1\rangle\langle 1| \Big)\nonumber\\
 &=\mathbb{P}_{\pi'}\Big(\rho_{_N}=|1\rangle\langle 1| \Big),
\end{eqnarray}
for any $\pi=\{\pi_k\}_{k=0}^{N-2}$, where $\pi'=\pi \oplus E_\ast=\{\pi_k\}_{k=0}^{N-1}$ with $\pi_{N-1}=E_\ast$. As a result,  the following relation holds between the optimal policies under the two objectives ${\mathrm{J}}_\pi (N)$ and $\tilde{\mathrm{J}}_\pi (N)$.

\begin{proposition}\label{prop2}
It holds that  $\max_\pi {\mathrm{J}}_\pi (N)= \max_\pi \tilde{\mathrm{J}}_\pi (N-1)$. In fact,  $\pi^{\star}(N)=\pi^\diamondsuit(N-1)\oplus E_\ast$ with  $E_\ast=\{|0\rangle \langle 0|, |1\rangle \langle 1| \}$.
\end{proposition}

The intuition behind Proposition  \ref{prop2} is that one would expect to get as closely as possible to  the target state at step $N-1$, if one tends to successfully project onto the target state at step $N$.  We also know from Proposition \ref{prop2} that we can solve the maximal expected fidelity problem in $N$ steps by the solutions of maximizing the arrival probability in $N+1$ steps.

Similarly, we can also find the optimal policy $\pi^\diamondsuit$ for the objective $\tilde{\mathrm{J}}_\pi (N)$ using dynamical programming.     Define the cost-to-go function $ \tilde{\mathbf{V}}(k,x)$ for  $\tilde{\mathrm{J}}_\pi (N)$ as
\begin{eqnarray}
 \tilde{\mathbf{V}}(k,x)=\max_\pi  \mathbb{E}_\pi \Big[\langle 1| \rho_{_N} |1\rangle\  \Big|\  \rho_k=x\Big]
\end{eqnarray}
for $ k=0,\dots,N$.
Then  $\tilde{\mathbf{V}}(k,x)$ satisfies the following  recursive equation
\begin{eqnarray}
\tilde{\mathbf{V}}(k,x)=\max_{u\in\mathcal{E}}\sum_{y\in \mathcal{Y}}\mathbb{P}\Big(y\Big| u,x\Big)\tilde{\mathbf{V}}\Big(k+1, \mathcal{M}_{u}^{y}(x)\Big),
\end{eqnarray}
for $k=0,\dots,N-1$,
with terminal condition
\begin{equation}
\tilde{\mathbf{V}}(N,x)=\tr\big(x |1\rangle \langle 1|\big).
\end{equation}
The optimal policy  $\pi^\diamondsuit$ can be obtained by solving
\begin{eqnarray}
\pi^\diamondsuit_k(\rho_k)=\arg \max_{u\in\mathcal{E}}\sum_{y\in \mathcal{Y}}\mathbb{P}\Big(y\Big|u,\rho_k\Big)\tilde{\mathbf{V}}\Big(k+1, \mathcal{M}_{u}^{y}(\rho_k)\Big) \nonumber
\end{eqnarray}
for $ k=0,\dots,N-1$. The maximal expected fidelity $\tilde{\mathrm{J}}_{\pi^\diamondsuit} (N)=\tilde{\mathbf{V}}(0,\rho_0)$.

\subsection{Minimal Arrival Time}
In previous discussions the deadline $N$ plays an important role in the objective functionals as well as  in their solutions. We now consider the case when the deadline is flexible, and we aim to  minimize  the average number of steps it takes to arrive at the target state. Now the control policy is denoted as $\pi=\{\pi_k\}_{k=0}^\infty$, where $\pi_k$ selects a measurement from the set $\mathcal{E}$. Associated with $\pi$, we define
\begin{eqnarray}
\mathscr{A}_\pi:= \inf_k\Big\{ \rho_k=|1\rangle \langle 1|\Big\}.
\end{eqnarray}
Note that  $\mathscr{A}_\pi$ defines a stopping time (cf., \cite{durett}) associated with the random processes $\{\rho_k\}_0^\infty$, and we assume that $\pi$ is {\it proper} in the sense that $$
\mathbb{P}_\pi\Big(\mathscr{A}_\pi<\infty \Big)=1.
$$
We continue to introduce
\begin{equation}
\mathrm{J}_\pi^\flat =\mathbb{E}_\pi \big[\mathscr{A}_\pi \big]
\end{equation}
as the objective functional, which is  the expected time it takes for the quantum state to reach the target $|1\rangle \langle 1|$  following policy $\pi$. Minimizing $\mathrm{J}_\pi^\flat$ is a {\it stochastic shortest path problem} \cite{Bertsekas-TsitsiklisMOR1991}.

We introduce $\mathscr{B}_\pi(x):=\inf_k\big \{ \rho_k=|1\rangle \langle 1| \ \big| \ \rho_0=x \big \}$ and
\begin{equation}
\mathbf{V}^\flat(x)= \min_\pi \mathbb{E}_\pi\Big[\mathscr{B}_\pi(x)  \Big].
\end{equation}
 The Markovian property of $\{\rho_k\}_{k=0}^\infty$ leads to that  the optimal policy $\pi^\natural$ is {\it stationary} in the sense that $\pi_k=\pi^\natural (x)$ for all $k$. The following conclusion holds applying directly the results of \cite{Bertsekas-TsitsiklisMOR1991}.

%

\begin{proposition}
The cost-to-go function $\mathbf{V}^\flat$ satisfies the following recursion
\begin{equation}
\mathbf{V}^\flat(x)=1+\min_{u\in\mathcal{E}}\sum_{y\in \mathcal{Y}}\mathbb{P}\Big(y\Big| u,x\Big) \mathbf{V}^\flat\Big(   \mathcal{M}_{u}^{y}(x) \Big),
\end{equation}
for all $x \neq |1\rangle \langle 1|$,
with boundary condition   $\mathbf{V}^\flat(|1\rangle \langle 1|)=0$. The optimal policy $\pi^\natural$  is  given by
\begin{equation}
\pi^\natural (x)=\arg \min_{u\in\mathcal{E}} \sum_{y\in \mathcal{Y}}\mathbb{P}\Big(y\Big| u,x\Big) \mathbf{V}^\flat\Big(   \mathcal{M}_{u}^{y}(x) \Big).
\end{equation}
The optimal $\mathrm{J}_{\pi^\natural }^\flat$ is given by $\mathrm{J}_{\pi^\natural }^\flat=\mathbf{V}^\flat(\rho_0)$.
\end{proposition}

Technically it cannot be guaranteed that for any given measurement set $\mathcal{E}$, there always exists at least one policy $\pi$ under which $\mathrm{J}_\pi^\flat$ admits a finite number. However, some straightforward calculations indicate that for the  set $\mathcal{E}$ of projective measurements given in Eq. (\ref{MeasurementSet}), finite $\mathrm{J}_\pi^\flat$ can always be achieved for a class of policies.

\begin{table*}
\begin{center}
\renewcommand\arraystretch{1.5}
\resizebox{10cm}{!} {
\begin{tabular}{|c|c|c|c|c|c|c|c|c|c|c|}
  \hline
   $x$& $|0\rangle$ & $|1\rangle$ & $|\phi_1\rangle$ & $|\psi_1\rangle$ & $|\phi_2\rangle$ & $|\psi_2\rangle$ & $|\phi_3\rangle$ & $|\psi_3\rangle$  &  $|\phi_4\rangle$ & $|\psi_4\rangle$ \\
  \hline
  $ \pi^\natural(x)$ & $E_2$ & $E_5$ & $E_3$ & $E_5$ & $E_4$ & $E_5$ & $E_5$ & $E_1$ & $E_5$ & $E_2$ \\
  \hline
  \end{tabular}
  }
  \caption{\rm The optimal policy $\pi^\natural$ minimizing the expected time it takes for the quantum state to reach the target state $|1\rangle\langle 1|$ for control set $\mathcal{E}_\ast$ with $T=5$.}
\label{OptimalStoppingTable}
\end{center}
\end{table*}

\subsection{Numerical Example: Minimal Arrival Time}
\begin{figure}[t]
\begin{center}
\includegraphics[width=10cm]{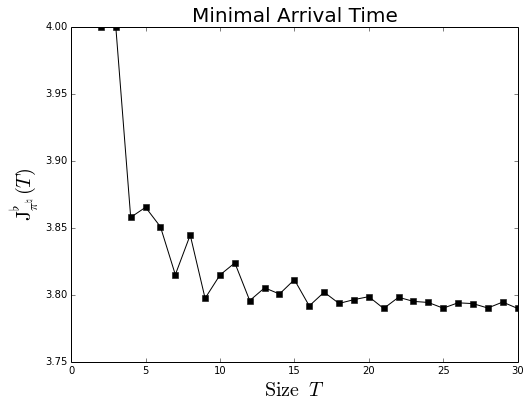}
\caption{The minimized average number of steps it takes to arrive at the target state $|1\rangle\langle 1|$ from the initial state $|0\rangle\langle 0|$ employing control set $\mathcal{E}_\ast$ of size $T$. }
\label{OptimalStopping}
\end{center}
\end{figure}

Again, consider  $T$   projective measurements from the set  \cite{pechen2006quantum}
$$
\mathcal{E}=\Big\{E_i,\ \  i=1,2,\ldots, T \Big\}.
$$
In Figure \ref{OptimalStopping}, we plot $\mathrm{J}_{\pi^\natural }^\flat(T)$ as a function of $T$, for $T=2,3,\ldots,30$. Numerical calculations show that the minimized average number of steps of driving $|0\rangle\langle 0|$ to $|1\rangle\langle 1|$ does not depend too much on the size of control set, it oscillates around $3.8$ for control sets of reasonable size. Also for measurement  set $\mathcal{E}_\ast$ with $T=5$, we show the optimal policy $\pi^\natural$ in Table \ref{OptimalStoppingTable}.

\section{Conclusions}\label{sec4}
We have  proposed   feedback designs  for  manipulating a quantum state to a target state by performing sequential   measurements. Making use of   Belavkin's quantum feedback control theory,  we showed that finding the measurement selection  policy  that   maximizes  the probability of successful state manipulation is an optimal control problem which can be solved by dynamical programming for any  given set of measurements and a given time horizon. Numerical examples  indicate that making use of  feedback information  significantly improves the success probability compared to classical scheme without taking feedback. It was shown that the probability of reaching the target state via feedback policy reaches  $0.9968$ using merely $10$ steps, while  classical results \cite{roa2007quantum,pechen2006quantum}  suggested that  naive strategy via consecutive measurements in turn reaches success probability one when the number of steps tends to infinity.  Maximizing the expected fidelity to the target state and  minimizing the expected arrival time were also considered, and some  connections and differences among these objectives were also discussed.

\medskip

\noindent {\bf Acknowledgments}

We gratefully acknowledge support  by the  Australian Research Council Centre of Excellence for Quantum Computation and Communication Technology (project number CE110001027), and AFOSR Grant FA2386-12-1-4075).

\end{document}